# Innovative Oxide Transistor Satisfying Performance and Reliability Simultaneously by Understanding of Physics and Materials Properties

C.W. Cheng, J. Smith, P. Solomon, R. Watters, D. Piatek, C. Lavoie, M. Hopstaken, L. Gignac, D. Bishop, B. Khan, M. BrightSky, G. Gionta, P. Hashemi, V. Narayanan, M. M. Frank
IBM T.J. Watson Research Center, Yorktown Heights, NY 10598 USA, email: chengwei@us.ibm.com

*Abstract*—— Guided by a comprehensive analysis of accumulation mode transistor physics and oxide semiconductor materials properties, we demonstrate an innovative oxide semiconductor transistor structure and process flow that breaks the constraint between performance and reliability observed in conventional InGaZnO$_4$ (IGZO) transistors. The newly proposed 10 nm IGZO transistor features high on current, high extrinsic mobility (20 cm$^2$V$^{-1}$s$^{-1}$), near-zero hysteresis, and only 15 mV V$_t$ shift after positive-bias-stress (PBS) of 3 MV/cm stress for 1000s at room temperature.

## Introduction & Oxide Transistor Physics

Silicon and oxide field-effect-transistors (FET) are inversion and accumulation mode devices, respectively. While Si FET in CMOS logic technology has a top gate and top source-drain (SD) contacts, oxide FET commonly employs a bottom gate, top SD contact staggered configuration [1]. Despite these differences, their electrical characteristics look comparable, and Si FET electrical models and concepts are thus often directly applied to oxide FET. Fig. 1 shows the relation between surface charge (|Qs|) and surface potential ($\Psi_s$) for IGZO and Si. Surface charge of an accumulation (inversion) channel increases exponentially with surface potential beyond flatband (threshold) and the channel charge Qs is then proportional to C$_{ox}$ (V$_g$-V$_{fb}$ (V$_t$)); note that the role of V$_{fb}$ in oxide FET is analogous to V$_t$ in Si FET [2]. While the physics behind inversion and accumulation layer in Si and oxide FET are in fact similar, important differences are often neglected. In fact, **there are at least two channels (accumulation layer and bulk film)** in oxide FET instead of a single inversion channel in Si FET due to the lack of p-n junction isolation in the SD regions (Fig. 2). Fig. 3 shows the band structure and Id-Vg curve of an oxide transistor in different operation regimes. While the threshold current in Si channels flows near the gate dielectric interface, **the threshold and bulk current in oxide FET start near the back-channel interface opposite to the gate side**. The channels of oxide FET are shut off by the depletion layer, where the depletion width is a function of carrier concentration, gate capacitance, and applied depletion gate voltage ($\Delta V_d$) (Fig. 3(c)). **Bulk current is dependent on channel carrier concentration, mobility, and oxide thickness, while accumulation charge and current are only gate voltage dependent.** Charge distribution and equivalent circuit of an oxide transistor are shown in Fig. 4.

## Materials and Device Fabrication

Shadow mask fabrication of long-channel transistors with sputter deposition process, 50 nm dry SiO$_2$ gate dielectrics, and p+Si global bottom gate were intentionally utilized to minimize any influence of hydrogen and processing damage during local bottom gate and device fabrication process, thus uncovering the intrinsic properties of the oxide materials. Details of channel and SD materials, fabrication process, and device structure are shown in Fig. 5.

## Oxide Material Properties

Understanding the fundamental material properties of oxide channel is important to develop appropriate fabrication process flow and thermal treatments. Fig. 6 shows the resistance-temperature (R-T) behavior and synchrotron X-ray diffraction of In$_2$O$_3$, IGZO, and ITO film deposited in different process conditions. Depositions with oxygen show higher as-deposited resistance. Mild plasma damage on the oxide would introduce surface damage and cause a highly conductive path on the surface. Oxygen annealing after deposition is effective to increase the resistance of In$_2$O$_3$ and IGZO by recovering surface damage and reducing oxygen vacancy, unlike for ITO due to crystallization and Sn doping.

Accordingly, FETs fabricated with the same process flow but different oxide semiconductors deliver different results. Fig. 7(a,b,c) shows Id-Vg curves of staggered In$_2$O$_3$, IGZO and ITO oxide transistors with ITO SD and SiO$_2$ encapsulation. The as-fabricated IGZO transistor before annealing shows weak modulation and low current, but it gradually improves after higher temperature annealing. In contrast, the In$_2$O$_3$ transistor shows shunting behavior, and gate modulation is recovered after higher temperature annealing due to the recovery of surface damage originating from ITO SD and SiO$_2$ deposition. The ITO transistor behaves similarly, but damage recovery is weaker and finally shunts again due to crystallization and Sn doping. This property makes ITO a good material for SD contact in oxide transistor. Process damage control (plasma, chemical, X-ray, etc.) throughout the integration flow is thus crucial to realize high performance and reliable oxide transistor.

To avoid any deposition plasma damage during SD or encapsulation deposition, a coplanar ITO transistor was fabricated without SiO$_2$ encapsulation. It shows decent gate modulation, and high current even in as-fabricated state without any thermal annealing (Fig. 7(d)). Electrical properties are further improved with low temperature annealing. However, this ITO transistor degrades when exposed to the full BEOL thermal budget (400$^o$C) and is thus not BEOL compatible. Reducing Sn doping might be a viable method to realize ITO transistor for BEOL applications.

## Thickness Dependent Electric Properties

Transistor performance (on current I$_{on}$) and reliability (hysteresis and PBS) are thickness dependent. Fig. 8 shows the maximum I$_{on}$ and Id-Vg hysteresis of IGZO transistors (without SiO$_2$ encapsulation, with oxide channel thickness ranging from 5 nm to 300 nm. The thinner the oxide channel, the better the performance but the worse the reliability.

Thickness-dependent I$_{on}$ is due to the extra access resistance (r$_{acss}$) caused by current flowing between the top SD contacts to the bottom accumulation channel (fig. 4). Fig. 9(a, b) shows the extracted contact resistivity ($\rho_c$) and sheet resistance (Rs) vs. gate bias and it shows clearly that only contact resistivity is thickness dependent and Rs is only gate bias dependent for IGZO thicker than 10 nm. The measured Rc is composed of interfacial contact resistance r$_c$ and r$_{acss}$. The contact resistance r$_c$ between SD contact and film only depends on the carrier concentration, while r$_{acss}$ also depends on oxide thickness. It is straightforward to think that r$_{acss}$ could be eliminated if a coplanar structure was adopted. However, the contact resistance in a coplanar structure is higher than that in a staggered structure (Fig. 9(c)), although gate voltage dependent Rs is the same in both structures. The coplanar structure features greater process simplicity and avoids material process damage, but it gives lower performance. To revitalize the coplanar structure, a modified coplanar process is proposed. Instead of depositing ITO SD contact as first step, a thin layer of channel material is deposited before the ITO SD contact and then oxide channel material is deposited between SD and partially onto the ITO. This is effective to boost the current to that of the staggered structure (Fig. 9d). **This modified coplanar structure further provides the freedom to tune or optimize the device performance by inserting high carrier concentration material only in the SD region to further reduce contact resistance, which** could not be achieved by traditional process flow due to the impact on sub-threshold slope and on/off ratio.

Apparent negative contact resistance and transfer length extracted from transmission line method (TLM) data are seen commonly on samples with very thin thickness or damage on the oxide semiconductor surface [5]. Fig. 10 shows TLM data from 5 nm IGZO devices, taken 10 times consecutively. The first measurement gives a normal TLM result, but 2Rc and transfer length gradually become negative at low gate voltages during repeated measurements, and Rs data gradually shift positively. Raw TLM data of 1st and 10th measurements are shown in fig. 10(d), confirming the observation of negative Rc. This is due to the voltage stress gradually accumulated in each measurement.

PBS and Negative Bias Stress (NBS) also show significant degradation for thinner channels (Fig. 11). Typically, NBS is much better than PBS for oxide transistors. It is also interesting to see that the NBS of a 10 nm IGZO transistor shows positive shift during test, and this is because the PBS is so bad for ultra-thin film that Id-Vg curve shifts more significantly even in short Id-Vg curve measurement period than negatively in negative bias stress period.

It is speculated that PBS would cause a positively charged donor band tail states on the surface oxide channel, causing $V_t$ to shift positively and potentially giving rise to a third non-ideal channel to contribute current after threshold current and bulk channel are on. Fig. 12(a) shows the Id-Vg curves from a 5 nm IGZO device measured after PBS to 5, 10, and 15 V for 1000s consecutively. As described in the previous physics sections, current flow starts from the surface of the oxide channel in staggered structure. This PBS-induced damage layer would then contribute current immediately after the bulk channel is open. This would cause a much sharper transition from threshold to channel-on state in Id-Vg curves. As shown above, PBS is one of the root causes for apparent negative contact resistance in TLM. Any surface damage during deposition and processing could also cause a third non-ideal channel on the oxide semiconductor surface (Fig. 12(b)). These extra non-ideal channels would cause the channel shortening effect if there is only one channel considered in TLM, while also causing apparent negative contact resistance and transfer length. (Fig. 12(c))

Appropriate passivation and encapsulation of the oxide channel surface are the key to good reliability, in addition to the gate dielectrics/oxide channel interface. Fig. 11(d) shows the PBS and NBS results after $SiO_2$ encapsulation on 50 nm IGZO. The $SiO_2$ layer significantly reduces the $V_t$ shift from few volts to 50 mV in PBS and from -110 mV to -25 mV in NBS at room temperature (RT) compared to a device without $SiO_2$ encapsulation. However, the PBS results are still thickness dependent even with high-quality $SiO_2$ passivation. High performance devices need thinner oxide channels to reduce access resistance for higher current while high reliability devices need thicker oxide channels to have better PBS for long term stable operation. It is a dilemma and conflict between performance and reliability with thickness dependency.

## INNOVATIVE TRANSISTOR AND PROCESS

**An innovative device structure and process flows for high performance and reliability were demonstrated to break the constraint to satisfy different demands on the thickness in channel and contact region.** Fig.13(a) shows this innovative structure achieved by "Addition" and "Subtraction" processes. Addition process follows the standard staggered device fabrication process flow but deposits more channel material on top of the channel region before encapsulation, while Subtraction process starts with thicker oxide channel material with encapsulation, and then the contact region is etched to leave a thin oxide semiconductor layer for SD contact. Subtraction process needs less steps, but it requires more attention to manage the etching damage in SD contact area. **Devices data** from both fabrication processes **show excellent on current with zero hysteresis** (Fig. 13(b,c)). Fig. 13(d) shows the comparison of hysteresis data of conventional 5, 10, and 100 nm IGZO channel and innovative 5 nm and 10 nm IGZO channel contact device data. Maximum on current is higher for thinner channel contact in the innovative structure.

High extrinsic mobility (20 cm$^2$V$^{-1}$s$^{-1}$) and good PBS (15 mV) were demonstrated by an innovative IGZO transistor with 10 nm channel contact from Addition process flow. Fig. 14(a,b,c) shows the mobility extracted from the saturation region, mobility versus accumulation channel charge, and the PBS result. By adopting a thin channel contact SD region, the Rc contribution could be reduced substantially, and the extrinsic mobility approaches the intrinsic IGZO mobility. Indeed, an innovative IGZO transistor with 5 nm channel contact even gives 22 cm$^2$V$^{-1}$s$^{-1}$ extrinsic mobility due to even lower $r_{acss}$. The $V_t$ shift at RT shows a huge improvement, from a few volts for an IGZO control device to 15 mV for an innovative IGZO transistor at 10 nm IGZO channel contact thickness. (Fig. 14(d))

In conclusion, the physics behind oxide transistors was reviewed. **A multi-channel concept incorporating ideal and non-ideal effects needs to be adopted to interpret transistor and TLM data correctly.** Fundamental materials properties were comprehensively explored and studied. A conflicting thickness dependence on transistor performance and reliability was found. **An innovative transistor structure and process flow were proposed to break this constraint and to satisfy performance and reliability simultaneously.** Appropriate passivation and encapsulation on oxide channel surface was proven to be very important for reliability improvement.

We thus provide a guideline and give more freedom to further optimize oxide transistors with higher performance and better reliability for higher definition display by reducing pixel size, and to help realize oxide semiconductor applications in the BEOL of VLSI circuits.

## ACKNOWLEDGEMENTS

The authors are grateful for support from Microelectronic Research Laboratory, Dr. N. Gong, Dr. T.C. Chen, and Dr. H. Bu at IBM Research and Synchrotron XRD from Canadian Light Source.

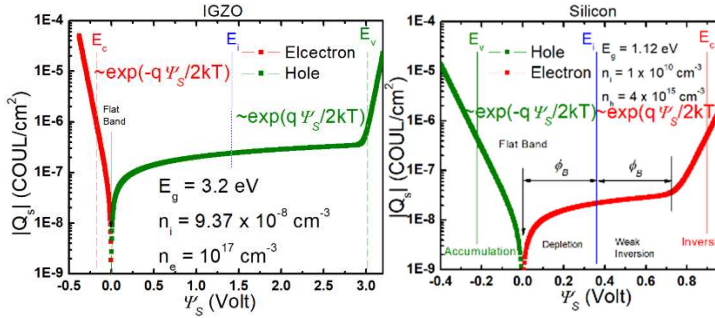
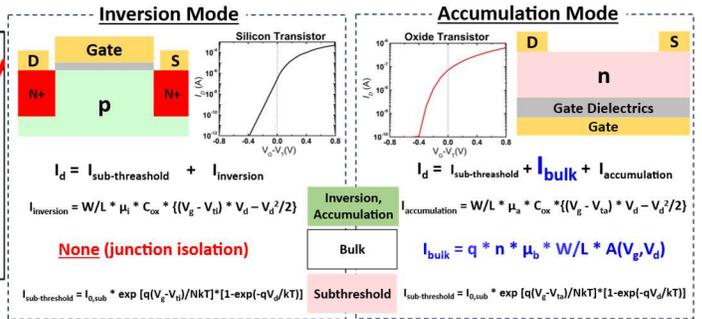

Fig 1. Surface charge ($Q_s$) versus surface potential ($\Psi_s$) of IGZO and Silicon.

Fig. 2. Structure, transfer characteristics, and electric equations of inversion (Silicon) and accumulation (Oxide) transistor [3,4]

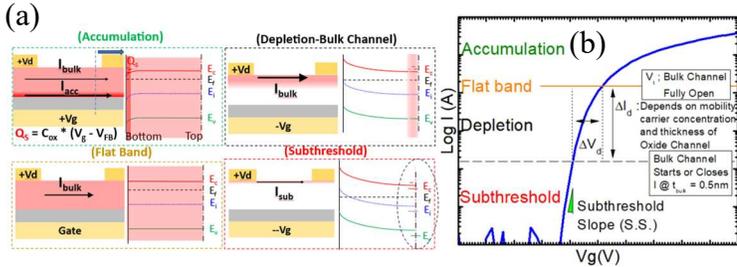
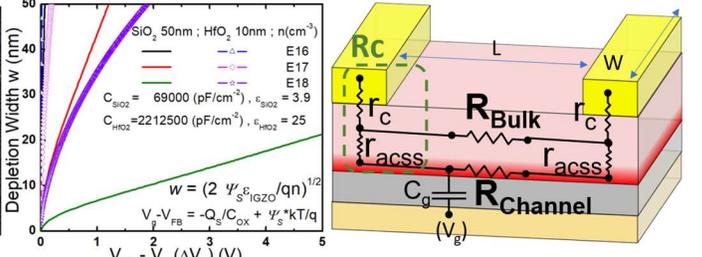

Fig. 3. (a) Band structure and current flows inside oxide transistor in different operation regimes. (b) Id-Vg curve of oxide transistor with defined operation regimes (c) Calculated depletion gate voltage ($\Delta V_d$) vs. depletion width (w) of carrier concentrations with 50nm $SiO_2$ and 10 nm $HfO_2$ as gate dielectrics. Color depth represents carrier concentration.

Fig. 4. Equivalent circuit of staggered oxide transistor. Accumulation and bulk channels exist simultaneously

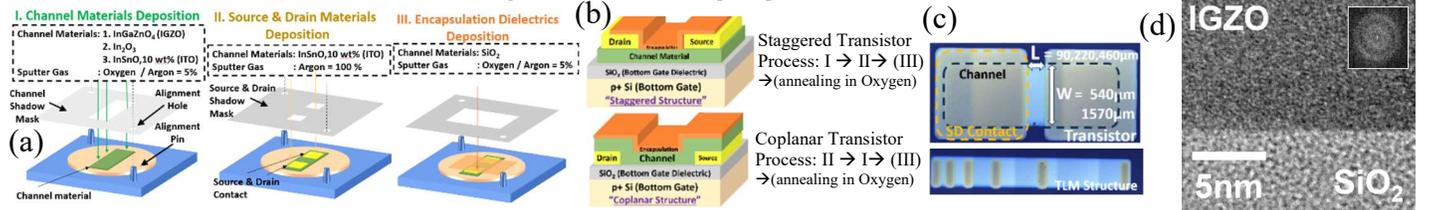

Fig. 5. (a) Shadow mask device fabrication process. Shadow mask alignment setup is precisely designed and manufactured to ensure good alignment of layers (b) Staggered and coplanar transistor structure and process flow with shadow masks (c) Photo of fabricated transistor and TLM (Transmission Line Method) structure (d) TEM image of IGZO/$SiO_2$ interface. Inset: Diffraction pattern of IGZO. IGZO is amorphous

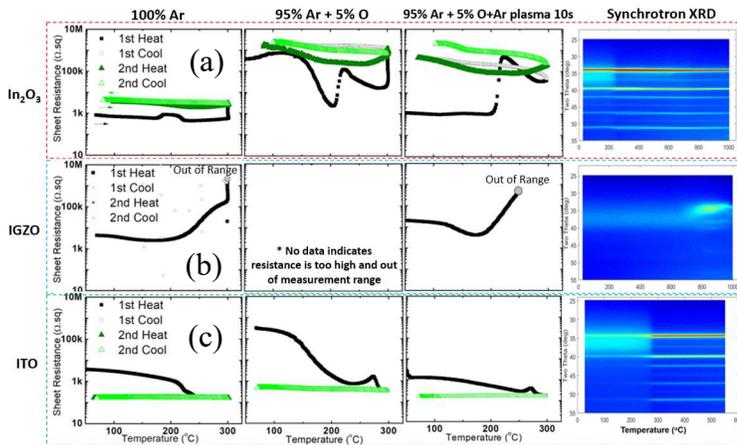
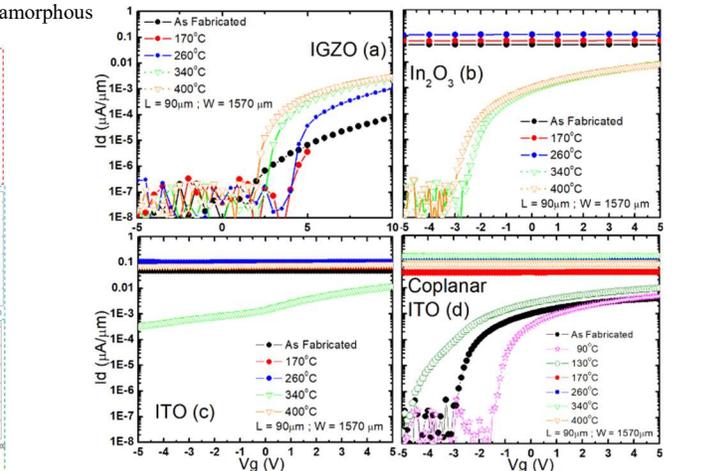

Fig. 6. Sheet Resistance (Rs) and Synchrotron XRD versus temperature (T) measurements of 25nm (a) $In_2O_3$ (b) IGZO and (c) ITO sputter deposited with pure argon, 5% oxygen in argon, and short argon plasma damage after deposition. Rs-T curves were measured in oxygen atmosphere twice continuously.

Fig. 7. Id-Vg curves of 30nm $SiO_2$ encapsulated staggered 5nm (a) IGZO (b) $In_2O_3$ (c) ITO and (d) non-$SiO_2$ encapsulated coplanar ITO transistors with sequential temperature ramping annealing in oxygen.

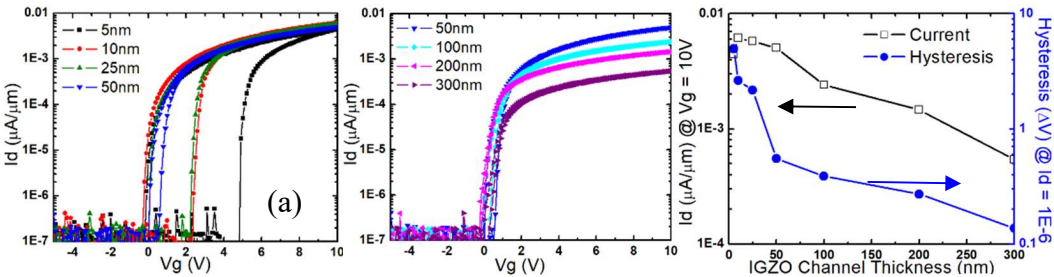

Fig. 8. Hysteresis Id-Vg curves of non-$SiO_2$ encapsulated IGZO transistor with channel thickness ranging (a) 5-50nm and (b) 50-300nm. (c) Summary of maximum on currents and hysteresis of devices from (a) and (b). **Thickness dependent performance and reliability are observed.**

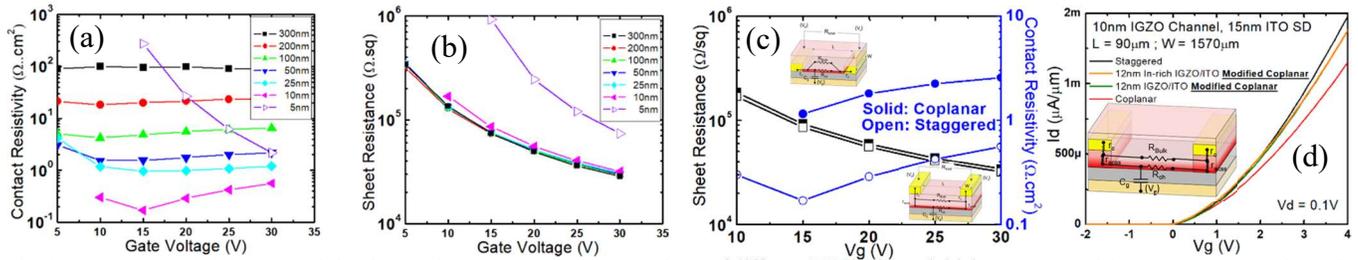

Fig. 9. (a) Contact resistivity ($\rho_c$) and (b) Sheet resistance (Rs) versus gate voltage of different IGZO channel thickness extracted from TLM. (c) $\rho_c$ and Rs of 10nm IGZO TLM results extracted from staggered and coplanar structures (d) Linear Id-Vg curve of modified coplanar contact and device structure

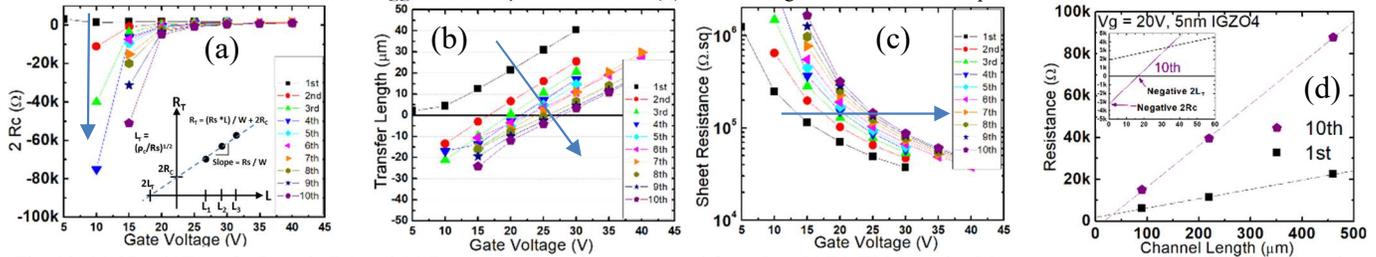

Fig. 10. (a) 2Rc (b)Transfer Length (Lt) and (c) Rs versus gate voltage extracted from 5nm IGZO TLM results. Measurements were taken for 10 consecutive times. (d) TLM results from 1st and 10th measurements @ Vg = 20V. Inset figures in (a) example of TLM method (d) Zoom-in of 2Rc and 2Lt intersections

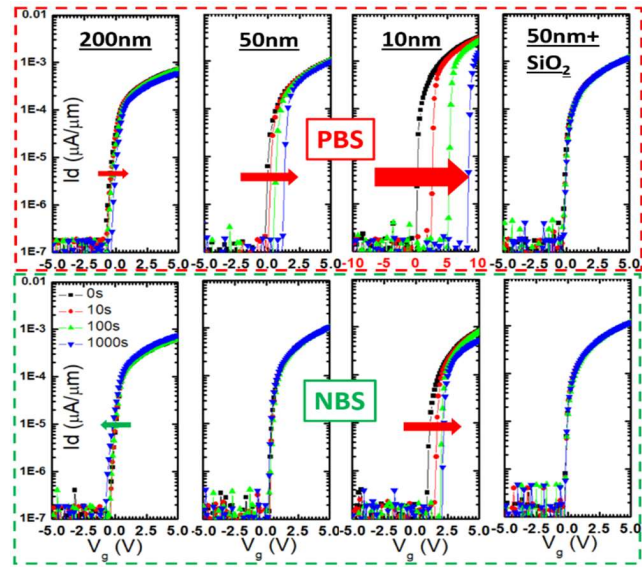

Fig. 11. PBS and NBS results of (a) 200 (b) 50 and (c) 10nm IGZO transistors without SiO$_2$ encapsulation and (d) 50nm IGZO transistor with 30nm SiO$_2$ encapsulation (W/L =220/1570µm). 3MV/cm gate voltage stress were used for PBS (+15V) and NBS (-15V).

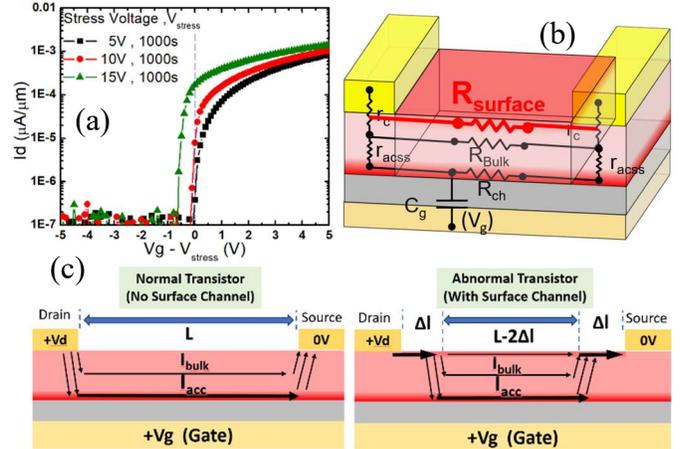

Fig. 12. (a) Id-Vg curves of 5nm IGZO transistor after 5,10, and 15V gate voltage stress for 1000s. X-axis were shifted by V$_{stress}$. (b) Equivalent circuit of oxide transistor with existence of 3rd non-ideal leaky channel on oxide surface (c) Explanation of channel shortening effect commonly observed on oxide transistors. Channel shortening value Δl is influenced by surface channel conductivity, gate voltage, carrier concentration and thickness of oxide channel

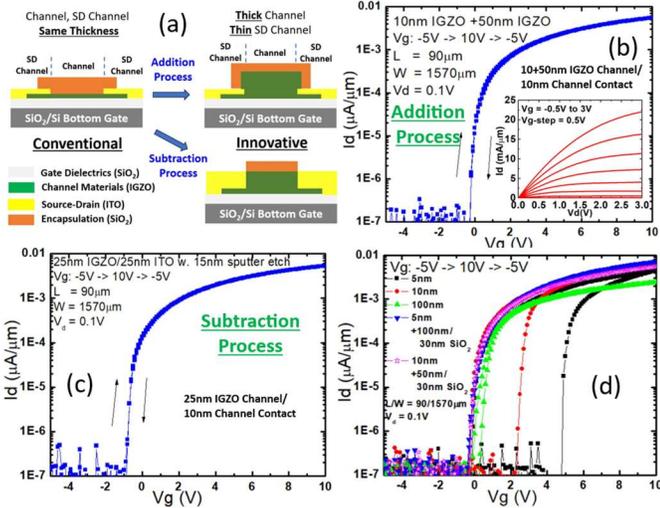

Fig. 13. (a) Innovative Oxide Transistor and process flows. (b, c) Hysteresis of Id-Vg curves of 10nm IGZO channel contact from Addition and Subtraction process. (d) Summary of hysteresis of Id-Vg curves of regular and innovative 5 and 10nm IGZO transistors. Zero hysteresis observed from all innovative oxide transistors. Inset in (b) Output characteristics

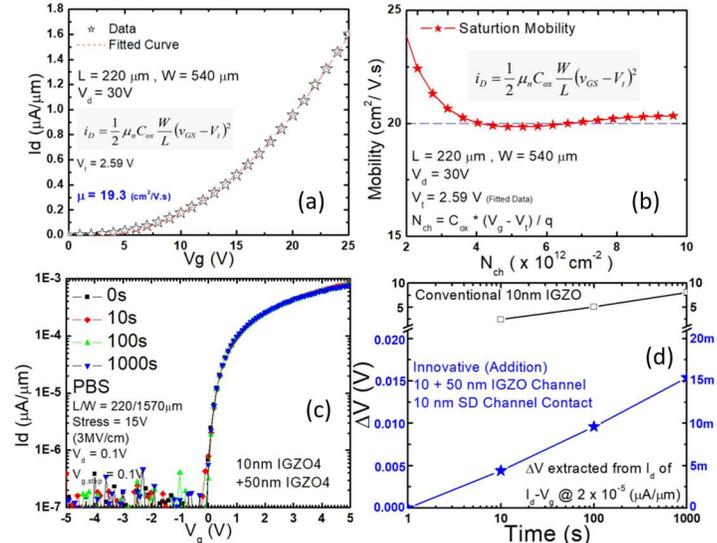

Fig. 14. (a) Saturation Id-Vg curve for mobility extraction. (b) mobility versus accumulation channel charge. (c) PBS of 10nm innovative IGZO transistor with 50nm IGZO channel capping from **Addition** process. (d) Comparison of V$_t$ shift of PBS results from conventional and innovative 10nm IGZO transistors. Only **15mV** V$_t$ shift after 3MV/cm 1000s stress at room temperature from innovative transistor with extrinsic mobility **20 cm$^2$/V.s**.